\documentclass[preprint,amsmath,amssymb,prd]{revtex4-1}
\usepackage{epsf}
\usepackage{dcolumn}
\usepackage{amsmath}
\usepackage{amssymb}
\usepackage{color}
\usepackage{bm}
\everymath{\displaystyle}

\usepackage[breaklinks=true,colorlinks=true,backref,pagebackref]{hyperref}

\begin{document}

\title{Dynamics of a tensor polarization of particles and nuclei and its influence on the spin motion in external fields}

\author{Alexander J. Silenko\footnote{Email: alsilenko@mail.ru}}

\affiliation{Bogoliubov Laboratory of Theoretical Physics, Joint Institute for Nuclear Research,
Dubna 141980, Russia \\
and Research Institute for Nuclear Problems, Belarusian State University, Minsk 220030, Belarus}


\begin{abstract} 
The tensor polarization of particles and nuclei is constant in a coordinate system rotating with the same angular velocity as the spin. In the laboratory frame, it rotates with this angular velocity. The general equation defining the evolution of the tensor polarization is derived. An explicit form of the dynamics of this polarization is given in the case when the angular velocity of the spin precession is constant and vertical. It is shown that the spin tensor interactions result in mutual transformations of the vector and tensor polarizations. These interactions can influence the spin motion.
\end{abstract}

\maketitle

\section{Introduction}\label{Introduction}

Polarized beams are often considered as a research tool to
study the fundamental interactions and to search for new physics. Unlike spin-1/2 fermions, particles and nuclei with spin $s\ge1$ possess a tensor polarization.
This polarization is an important property of particles and nuclei and can be measured with a good accuracy. Investigations of dynamics of tensor polarization of light nuclei (e.g., the deuteron with spin 1) allow one to predict and to describe some new effects \cite{Bar3,Baryshevsky,PRC2007,PRC2008,PRC2009,JPhysConfSer,BarHE,TechPhysLett}. These effects are conditioned either by quadratic in spin interactions of deuteron \cite{Bar3,Baryshevsky,PRC2007,PRC2008,PRC2009,JPhysConfSer,BarHE} or by quantum beats in positronium \cite{TechPhysLett}. To get a basic understanding of polarization effects, it is necessary to consider \emph{linear} in spin interactions of \emph{tensor-polarized} particles and nuclei. The dynamics of the tensor polarization in external fields has been described in Refs. \cite{EvTensorPolnLee,JPhysG2015}. In Ref. \cite{EvTensorPolnLee}, some examples of evolution of the tensor polarization have been investigated and some properties of a $5\times5$ spin transfer matrix defining this polarization have been considered. Evidently, the use of the $5\times5$ matrix is much more difficult than that of the corresponding $2\times2$ matrix for the vector polarization.

In the present work, we use a simple general approach for a description of dynamics of the tensor polarization of particles and nuclei when this dynamics is caused by linear in spin interactions with external fields. The approach proposed allows one to couple dynamics of the tensor and vector polarizations and to obtain general formulas defining an evolution of particles/nuclei polarization in external fields. We also show that the spin tensor interactions result in mutual transformations of the vector and tensor polarizations and can influence the spin motion.

The square and curly brackets, $[\dots,\dots]$ and $\{\dots,\dots\}$, denote commutators and anticommutators, respectively.


\section{General properties of spin dynamics}\label{dynamics}

As is known, the spin
rotation is exhaustively described with the polarization
vector $\bm P$ defined by
\begin{eqnarray}
P_i =\frac{<S_i>}{s}, ~~~ i=x,y,z. \label{eq1P}\end{eqnarray}
Here $S_i$ are corresponding spin matrices and $s$ is the spin
quantum number. Averages of spin operators are expressed by their convolutions with the wave function
$\Psi(t)$. Particles with spin $s\geq1$ also possess a tensor
polarization. Main characteristics of such a polarization are
specified by the polarization tensor $P_{ij}$, which is given by
\begin{eqnarray}
P_{ij} = \frac{3 <S_iS_j + S_jS_i>-2s(s+1)\delta_{ij}}{2s(2s -
1)}, ~~~ i,j=x,y,z. \label{eqPRC}\end{eqnarray} The polarization
tensor satisfies the conditions $P_{ij}=P_{ji}$ and
$P_{xx}+P_{yy}+P_{zz}=0$ and therefore has five independent
components. In the general case, the polarization vector and the
polarization tensor are time-dependent. Additional tensors
composed of products of three or more spin matrices are needed
only for the exhaustive description of polarization of
particles/nuclei with spin $s\ge3/2$.

The eight parameters defined by the polarization vector and the polarization tensor are independent.  In
particular, $<S_iS_j>\neq <S_i><S_j>$. 

The quantum-mechanical Hamiltonian describing the interaction of the spin with external fields contains 
terms linear and bilinear on the spin:
\begin{equation} {\cal H}=\bm\Omega\cdot\bm S+Q_{jk}S_jS_k.
\label{VveEtot} \end{equation} Here $\bm S$ is the matrix operator
describing the rest-frame spin. Since the spin matrices satisfy
the commutation relation
\begin{equation} [S_i,S_j]=ie_{ijk}S_k ~~~ (i,j,k=x,y,z),
\label{commuta} \end{equation} the first term in Eq.
(\ref{VveEtot}) defines the spin rotation with the angular
velocity $\bm\Omega$. The absolute value and the direction of
$\bm\Omega$ may arbitrarily depend on time.

Even when the second term in Eq. (\ref{VveEtot}) is omitted,
dynamics of the polarization tensor is not trivial. We cannot
characterize the particle spin wave function only by a single
frequency. For a spin-1 particle, there are two frequencies,
$+\Omega$ and $-\Omega$, defining three equidistant levels. For a
particle with the integer spin $s=N$, there are $2N$ frequencies,
$\pm\Omega$, $\pm2\Omega$, ..., $\pm N\Omega$, defining $2N+1$
equidistant levels. The single frequency of the spin rotation,
$\Omega$, originates from the specific commutation relations
(\ref{commuta}) for the spin operators $S_i$. For a spin-1
particle, this property of the spin operators manifests in the
fact that $(S_i)_{13}=(S_i)_{31}=0~(i=x,y,z)$. Therefore, the spin
operators mix only two neighboring levels. But it is not the case
for the tensor polarization operators. Three of them mix the
levels 1 and 3 because the components $(S_{x}^2)_
{13},~(S_{x}^2)_{31},~(S_{y}^2)_{13},~(S_{y}^2)_{31},~(S_{x}S_{y}+S_{y}S_{x})_{13}$,
and $(S_{x}S_{y}+S_{y}S_{x})_{31}$ are nonzero. As a result, the
frequency $2\Omega$ also appears. For particles with greater spins
($s\ge1$), the situation is still more complicated.

We consider electromagnetic interactions, namely,
coupling of spins to electric and magnetic fields.
The second term in Eq. (\ref{VveEtot}) characterizes the tensor
electric and magnetic polarizabilities and the quadrupole
interaction. It influences dynamics of both the polarization
vector and the polarization tensor. As a rule, this term is much
less than the first one. 
We can calculate the commutator $[{\cal H},S_i]$. 
The term under consideration
brings a correction of the order of $|Q_{jk}S_jS_k|$ to the
angular frequency of spin precession $\Omega$. The effect of this correction on the spin dynamics is rather complicate
and does not reduce to a change of $\bm\Omega$. While the effect is small, it cannot be neglected in some cases (see Refs.
\cite{Bar3,Baryshevsky,PRC2007,PRC2008,PRC2009}). Nevertheless, its influence can be significant only in special cases like deuteron EDM experiments or experiments specially designed to measure the tensor magnetic and electric polarizabilities. In Secs. \ref{dynamics}--\ref{VerticalAxis}, we will disregard the spin tensor interactions and will take into account only the first term in the interaction Hamiltonian (\ref{VveEtot}).

The equation of spin rotation is given by
\begin{equation} \frac{d\bm S}{dt}=\frac i\hbar[{\cal H},\bm S]=\bm\Omega\times\bm S.
\label{VveEfft} \end{equation} When the particle/nucleus moves in electromagnetic fields, $\bm\Omega$ is defined by the Thomas-Bargmann-Michel-Telegdi (T-BMT) equation extended on the electric dipole moment (EDM) (see Refs. \cite{FukuyamaSilenko,PhysScr} and references therein):
\begin{equation} \begin{array} {c}
\bm \Omega=-\frac{e}{mc}\left[\left(G+\frac{1}{\gamma}\right){\bm B}-\frac{\gamma G}{\gamma+1}({\bm\beta}\cdot{\bm B}){\bm\beta}-\left(G+\frac{1}{\gamma+1}\right)\bm\beta\times{\bm E}\right.\\
+\left.\frac{\eta}{2}\left({\bm E}-\frac{\gamma}{\gamma+1}(\bm\beta\cdot{\bm E})\bm\beta+\bm\beta\times {\bm B}\right)\right].
\end{array} \label{Nelsonh} \end{equation}
Here $G=(g-2)/2,~g=2mc\mu/(e\hbar s),~\eta=2mcd/(e\hbar s),~\bm\beta={\bf v}/c$, $\gamma$ is the Lorentz factor, and $\mu$ and $d$ are the magnetic and electric dipole moments.

The Pauli matrices together with the unit matrix describe the spin of a Dirac particle and generate an
irreducible representation of the \emph{SU}(2) group.
Algebraically, the \emph{SU}(2) group is a double cover of the
three-dimensional rotation group \emph{SO}(3). As a result, the spin dynamics defined by the Dirac equation fully corresponds to the classical picture of rotation of an intrinsic angular momentum (spin) in external fields. The angular velocity of spin rotation does not depend on the spin quantum number. Therefore, both the classical description and the formalism
based on the Pauli matrices are applicable to particles/nuclei with an
arbitrary spin if the effect of spin rotation is analyzed.
Certainly, this formalism becomes insufficient if one considers spin-tensor effects mentioned in Sec. \ref{Introduction}.

In Eq. (\ref{Nelsonh}), the terms dependent on the EDM are presented. An experimental search for the deuteron EDM should take into account an existence of the spin tensor interactions  \cite{Bar3,Baryshevsky,PRC2007,PRC2008,PRC2009,JPhysConfSer}.

\section{Evolution of the polarization tensor}\label{vectorpol}

Let us first consider the approach used by Huang, Lee, and Ratner \cite{EvTensorPolnLee} to describe an evolution of the tensor polarization.
Instead of the polarization tensor (\ref{eqPRC}), the following spin-tensor operators can be used \cite{EvTensorPolnLee}:
\begin{equation} \begin{array} {c}
T_0=\frac{1}{\sqrt2}\left(3S_z^2-2\right), ~~~T_{\pm1}=\pm\frac{\sqrt3}{2}\left(S_\pm S_3+S_3S_\pm\right), ~~~ T_{\pm2}=\frac{\sqrt3}{2}S_\pm^2,
\end{array} \label{TensrPolnLee} \end{equation} where $S_{\pm} = S_1\pm iS_2$.
To obtain observable quantities, these operators should be averaged.

The approach proposed in Ref. \cite{EvTensorPolnLee} is based on the T-BMT equation, uses the Frenet-Serret curvilinear coordinates, and disregards the tensor interactions in the Hamiltonian (\ref{VveEtot}). Expressing the spin vector in terms of its components,
\begin{equation} \begin{array} {c}
\bm S=S_1 \bm e_1+S_2 \bm e_2+S_3 \bm e_3,
\end{array} \label{Lee} \end{equation}
one obtains the following equation of spin motion:
\begin{equation} \begin{array} {c}
\frac{dS_{\pm}}{d\phi}=\pm iG\gamma S_{\pm}\pm iF_\pm S_3, ~~~ \frac{dS_{3}}{d\phi}=\frac i2\left(F_- S_+- F_+ S_-\right),
\end{array} \label{Leet} \end{equation}
where $F_\pm$ characterize the spin
depolarization kick and $\phi$ is the azimuthal orbit rotation angle. The spin tune, $G\gamma$, is the number of spin revolutions per orbit turn.

Equations (\ref{TensrPolnLee})--(\ref{Leet}) allow one to obtain the set of five equations defining the evolution of the tensor polarization.  These equations can be presented in the matrix form with a 5$\times$5 matrix \cite{EvTensorPolnLee}. Evidently, the use of $5\times5$ matrices instead of standard $2\times2$ accelerator matrices needs much more cumbersome derivations. Moreover, even investigations of spin-tensor effects performed in Refs. \cite{PRC2007,PRC2008,PRC2009} used three-component wave functions and $3\times3$ matrices. Another remark is an absence of a direct connection between evolutions of the polarization vector and the polarization tensor, while spin tensor interactions are not taken into consideration in Ref. \cite{EvTensorPolnLee}.
We propose a different approach which couples dynamics of
the tensor and vector polarizations and simplifies a description of the evolution of the
tensor polarization of particles/nuclei in external fields.

It is not evident whether the evolution of the polarization tensor also reduces to the rotation. To clear the problem, we use the following approach. Let us consider the Cartesian coordinate system rotating about the $z$ axis with the same angular velocity $\bm\Omega(t)$ as the spin. In the general case, this angular velocity depends on time. We may superpose the rotating and nonrotating coordinate systems at the initial moment of time $t=0$ ($\bm e'_i(0)=\bm e_i$). The rotating coordinate system is denoted by primes. 


The spin components in the rotating coordinate system remain unchanged:
\begin{equation} \frac{dS'_i}{dt}=0~~~(i=x,y,z).
\label{reunc} \end{equation}

As a result, all tensor polarization operators and all components of the polarization tensor are also unchanged in this coordinate system ($S'_iS'_j+S'_jS'_i=const,~P'_{ij}=const$). This important property shows that the tensor polarization of particles/nuclei with spin $s\ge1$ rotates in external fields similarly to the vector polarization. This is valid not only for electromagnetic interaction but also for other (weak, gravitational) interactions. Other products of the spin operators $S'_iS'_j\dots S'_k$ are also conserved in the rotating coordinate system.

To determine the particle polarization in the nonrotating coordinate system (lab frame), we need to connect the directions of basic vectors of the primed and unprimed coordinate systems. This problem can be easily solved. We choose the initial moment of time $t=0$. Evolution of any basic vector of the primed coordinate system coincides with that of the spin when it is initially directed along the considered basic vector. Therefore, the final directions of the three primed basic vectors are defined by the three final directions of the spin when its corresponding initial ones are parallel to the three Cartesian axes of the unprimed coordinate system.
As a result, dynamics of the polarization tensor reduces to that of the polarization vector. In particular, the evolution of the polarization tensor in accelerators and storage rings can be unambiguously defined with usual $2\times2$ spin transfer matrices. Thus, one need not apply much more cumbersome $5\times5$ matrices.

Time dependence of the polarization tensor defined in the lab frame can be easily expressed in terms of the basic vectors $\bm e'_i$. 
Since $\bm e'_i(0)=\bm e_i$ and
\begin{eqnarray}
S'_i(t)=\sum_{k}{(\bm e'_i(t)\cdot\bm e_k)S_k},
\label{eqppnn}\end{eqnarray} the polarization tensor is given by
\begin{eqnarray}
P_{ij}(t) = \frac{3}{2s(2s - 1)}\sum_{k,l}{\left[(\bm e'_i(t)\cdot\bm e_k)(\bm e'_j(t)\cdot\bm e_l) <S_kS_l + S_lS_k>\right]}- \frac{s+1}{2s - 1}\delta_{ij}. \label{peqptfn}\end{eqnarray}

This simple equation defines general dynamics of the tensor polarization of particles and nuclei in
external fields.

\section{Spin rotation about the vertical axis}\label{VerticalAxis}

As an example, let us consider the simplest case when the angular velocity of the spin precession is constant and vertical ($\bm\Omega=\Omega\bm e_z$). We use the cylindrical coordinate system which is rather convenient for a description of spin dynamics in storage rings (see Refs. \cite{RPJSTAB,PhysPartNuclLett2015}). The spin motion in accelerators and storage rings is defined relative to either the momentum direction \cite{EvTensorPolnLee} or the horizontal axes of the cylindrical coordinate system. 

When the beam is vector-polarized, the
direction of its initial polarization can be defined by the spherical angles
$\theta$ and $\psi$. In this case, $\theta=const$ and the azimuth $\psi$ rotates with the angular frequency $\Omega$ ($\psi=\psi_0+\Omega t$). We should mention that the angle $\psi$ is defined relative to the axes $\bm e_\rho$ and $\bm e_\phi$ but not relative to the axes $\bm e_x$ and $\bm e_y$. In particular, $\psi=0$ means that the spin is radially directed. In this case, the polarization vector and the polarization tensor are defined by (see, e.g., Refs. \cite{PRC2007,PRC2008,PRC2009})
\begin{equation}\begin{array}{c}
P_{\rho}(0)=\sin{\theta}\cos{\psi}, \qquad
P_{\phi}(0)=\sin{\theta}\sin{\psi}, \qquad P_{z}(0)=\cos{\theta}, \end{array} \label{eq4vc} \end{equation} \begin{equation}\begin{array}{c}
P_{\rho\rho}(0)=\frac32\sin^2{\theta}\cos^2{\psi}-\frac12,\quad
P_{\phi\phi}(0)=\frac32\sin^2{\theta}\sin^2{\psi}-\frac12,\quad
P_{zz}(0)=\frac32\cos^2{\theta}-\frac12,\\
P_{\rho\phi}(0)=\frac34\sin^2{\theta}\sin{(2\psi)},\quad
P_{\rho z}(0)=\frac34\sin{(2\theta)}\cos{\psi},\quad
P_{\phi z}(0)=\frac34\sin{(2\theta)}\sin{\psi}.
\end{array} \label{eq4tn} \end{equation}
 
When the direction of an initial \emph{tensor} polarization of a tensor-polarized beam obtained by mixing two incoherent beams with opposite vector polarizations is defined by the spherical angles $\theta$ and $\psi$,
$\bm P=0$ and the components of the polarization tensor are the same.

For spin-1 particles and nuclei, it is also convenient to consider the tensor-polarized beam with the zero
projection of the spin onto the preferential direction defined by the spherical angles $\theta$ and $\psi$. In this case, 
$\bm P=0$ and the tensor polarization is given by
\begin{equation}\begin{array}{c}
P_{\rho\rho}(0)=1-3\sin^2{\theta}\cos^2{\psi}, \quad
P_{\phi\phi}(0)=1-3\sin^2{\theta}\sin^2{\psi}, \quad
P_{zz}(0)=1-3\cos^2{\theta}, \\ 
P_{\rho\phi}(0)=-\frac32\sin^2{\theta}\sin{(2\psi)}, \quad P_{\rho z}(0)=-\frac32\sin{(2\theta)}\cos{\psi}, \quad P_{\phi
z}(0)=-\frac32\sin{(2\theta)}\sin{\psi}.
\end{array}\label{intvi}
\end{equation}

\section{Spin evolution conditioned by the tensor interactions}

It has been ascertained in Refs. \cite{Bar3,Baryshevsky,PRC2007,PRC2008,PRC2009,JPhysConfSer} that the tensor interactions lead to mutual transformations between the vector and tensor polarizations. We can show that this property is general.
An analysis can be made on the basis of spin-1 particle matrices. As is known, main properties of spin 
dynamics are common for particles with any spins. To consider the spin dynamics, we need to use basic properties of the spin matrices defined by Eq. (\ref{commuta}) and (for spin-1 particles) by the equation
\begin{equation}\begin{array}{c}
S_iS_jS_k+S_kS_jS_i=\delta_{ij}S_k+\delta_{jk}S_i.
\end{array}\label{intvi}
\end{equation}

These properties result in the following commutation relations:
\begin{equation}\begin{array}{c}
[S_\rho^2,S_\phi]=i\{S_\rho,S_z\},\quad [S_\rho^2,S_z]=-i\{S_\rho,S_\phi\},\quad
[S_\phi^2,S_\rho]=-i\{S_\phi,S_z\},\\ \,
[S^2_\phi,S_z]=i\{S_\rho,S_\phi\}, \quad
[\{S_\rho,S_\phi\},S_\rho]=-i\{S_\rho,S_z\},\quad [\{S_\rho,S_z\},S_\rho]=i\{S_\rho,S_y\},\\ \,
[\{S_\phi,S_z\},S_\rho]=2i(S^2_\phi-S^2_z),\quad [\{S_\rho,S_\phi\},S_z]=2i(S^2_\rho-S^2_\phi),\\ \,
[\{S_\phi,S_z\},S_\rho]=2i(S^2_\phi-S^2_z),
\end{array}\label{eqone}\end{equation}
\begin{equation}\begin{array}{c}
 [\{S_\rho, S_\phi\},S_\rho^2]=-iS_z,\quad
[\{S_\rho,S_z\},S_\rho^2]=iS_\phi,\\ \, [\{S_\phi,S_z\},S_\rho^2]=0,\quad
[S_\rho^2,S_\phi^2]=[S_\rho^2,S_z^2]=[S_\phi^2,S_z^2]=0.
\end{array}\label{eqtwo}\end{equation}
Other commutation relations can be obtained from Eqs. (\ref{eqone}) and (\ref{eqtwo}) by cyclic permutations of indices.

Now, unlike Eq. (\ref{VveEfft}), we take into account the tensor interactions in the term $(i/\hbar)[{\cal H},\bm S]$. Equations
(\ref{eqone}) and (\ref{eqtwo}) show that these interactions transform the vector polarization to the tensor one and the other way round, respectively. The both equations demonstrate that the tensor interactions do not lead to an additional rotation of the vector and tensor polarizations. 

Nevertheless, a \emph{seeming} change of the angular velocity of the spin rotation can take place. Any vector-polarized beam is also tensor-polarized. The part of the operator $d\bm S/(dt)$ conditioned by the tensor interactions also changes the pseudovector $\bm S$. This effect can be mistaken for the above-mentioned change of $\bm\Omega$. The effect is rather important for the deuteron EDM experiment (see Refs. \cite{Bar3,Baryshevsky,PRC2007,PRC2008,PRC2009,JPhysConfSer}).
Thus, the tensor interactions can influence the spin motion.

\textbf{Acknowledgements.}
This work was supported in part by the Belarusian Republican Foundation for Fundamental Research
(Grant No. $\Phi$18D-002) and
by the Heisenberg-Landau program of the Federal Ministry of Education and Research (Germany).


\end{document}